\documentstyle[11pt,epsfig]{article}
\title{\bf Lorentz violation in brane cosmology, accelerated expansion
and fundamental constants}
\author{F. Ahmadi\thanks{email: fa-ahmadi@sbu.ac.ir},
S. Jalalzadeh\thanks{email: s-jalalzadeh@sbu.ac.ir} and H. R.
Sepangi\thanks{email: hr-sepangi@sbu.ac.ir}
\\ {\small Department of Physics, Shahid Beheshti University, Evin,
Tehran 19839, Iran}}
\begin{document}
\maketitle

\begin{abstract}
The notion of Lorentz violation in four dimensions is extended to
a 5-dimensional brane-world scenario by utilizing a dynamical
vector field assumed to point in the bulk direction, with Lorentz
invariance holding on the brane.  The cosmological consequences of
this theory consisting of the time variation in the gravitational
coupling $G$ and cosmological term $\Lambda_4$ are explored. The
brane evolution is addressed by studying the generalized Friedmann
and Raychaudhuri equations. The behavior of the expansion scale
factor is then considered for different possible scenarios where
the bulk cosmological constant is zero, positive or negative.
\end{abstract}

\section{Introduction}
The possibility that Lorentz invariance may be violated at high
energies in $4D$ with testable consequence has become a subject of
much interest in the past few years \cite{MVJLM}. Tentative
results from quantum gravity and string theory point to a ground
state that may not be Lorentz invariant \cite{TJ}. String theory
also predicts that we may live in a universe with non-commutative
coordinates \cite{CDS} leading to violation of Lorentz invariance
\cite{CHKLO}. In addition, astrophysical observations point to the
presence of high energy cosmic rays about the
Greisen-Zatsepin-Kuzmin cutoff \cite{ZK}, results which may be
explained by a breaking down of Lorentz invariance
\cite{CK}-\cite{SS}. Moreover, due to the unboundedness of the
boost parameter, exact Lorentz invariance, while mathematically
elegant, is unverifiable and therefore suspect. Most works
exploring possible Lorentz violation have focused on non
gravitational physics, {\it i.e.} flat space-times. Much less has
been done to investigate Lorentz breaking in curved space-times.
In flat space-times, Lorentz violation is described by couplings
to constant symmetry breaking tensors $V_{a},W_{ab}$ and so on. To
formulate Lorentz breaking in a curved space-time without
destroying general covariance such tensors must become dynamical
tensor fields that satisfy effective field equations.

A straightforward method of implementing local Lorentz violation
in a gravitational setting is to imagine the existence of a tensor
field with a non-vanishing expectation value and couple this
tensor to gravity or matter fields. The simplest example of this
approach is to consider a single time-like vector field with fixed
norm. A special case of this theory was first introduced as a
mechanism for Lorentz violation by Kostelecky and Samuel
\cite{KS}.  In a different context, Bekenstein has proposed a
theory of gravity with a fixed-norm vector in order to mimic the
effects of dark matter \cite{JB}. Also, studies of vector fields
in a cosmological setting without the fixed norm have been done in
\cite{NW}-\cite{VK}. This vector field picks out a preferred frame
at each point in space-time and any matter field coupled to it
will experience a violation of local Lorentz invariance
\cite{DCVK,CF}.

General relativity cannot describe gravity at high enough energies
and must be replaced by quantum gravity theory. The physics
responsible for making a sensible quantum theory of gravity is
revealed only at the Planck scale. This cut-off scale marks the
point where our old description of nature breaks down and it is
not inconceivable that one of the victims of this break down is
Lorentz invariance. It is thus interesting to test the robustness
of this symmetry at the highest energy scales \cite{AP,CG,ABGG}.
As usual in high energy physics, if the scale characterizing new
physics is too high then it cannot be reached directly in collider
experiments. In this case cosmology is the only place where the
effects of new physics can be indirectly observed. Brane-world
models offer a phenomenological way to test some of the novel
predictions and corrections to general relativity that are implied
by M-theory. Lorentz violating effects have also been studied
within the context of the brane-world scenarios. In such models,
the space-time globally violates $4D$ Lorentz invariance, leading
to apparent violations of Lorentz invariance from the brane
observer's point of view due to bulk gravity effects. These
effects are restricted to the gravity sector of the effective
theory while the well measured Lorentz invariance of particle
physics remains unaffected in these scenarios \cite{csaba,
burgess}. In a similar vein, Lorentz invariance violation has been
employed to shed some light on the possibility of signals
travelling along the extra dimension outside our visible universe
\cite{stocia}. In a different approach a brane-world toy model has
been introduced \cite{rubakov} in an inflating $5D$ brane-world
setup with violation of $4D$ Lorentz invariance at an energy scale
$k$.

In this paper we consider Lorentz violation in a brane-world
scenario where it occurs in the bulk space along the extra
dimension. This is complementary to the above mentioned brane
world scenarios in that the Lorentz violation affects the bulk
space rather than the $4D$ brane. To this end, we consider the
theory suggested by Jacobson and Mattingly \cite{JM,kostel} where
the gravitational effects of the vector fields are described in
four dimensions. We generalize the theory to include higher
dimensional gravity and in particular the brane-world scenarios.
As is well known, brane-worlds are often studied within the
framework of the $5D$ Einstein field equations projected onto the
$4D$ brane, a prime example of which is the formulation of
Shiromizu, Maeda and Sasaki (SMS) \cite{SMS}. To study Lorentz
violation, we consider a vector $N^{A}$ along the extra dimension.
In doing so a local frame at a point in space-time is inevitably
selected as the preferred frame. Put in other words, the existence
of the brane defines a preferred direction in the bulk. We then
move on to study the effects of local Lorentz violation on the
dynamics of the brane. We assume no coupling between the matter
fields and the vector field as the brane observer does not feel
the presence of the preferred frame. This additional field
modifies the $4D$ Einstein equations with cosmological
implications which we investigate by studying the resulting
Friedmann and Raychaudhuri equations on the brane. We find that
Lorentz violation allows for the construction of models which show
variations in the fundamental physical ``constants.'' Finally, we
determine the brane evolution for different possible scenarios
where the bulk cosmological constant is negative, zero or
positive.
\section{Field Equations}
In the usual brane-world scenarios the space-time is identified
with a singular hypersurface (or 3-brane) embedded in a
five-dimensional bulk. Suppose now that $N^{A}$ is a given vector
field along the extra dimension, effectively making the associated
frame a preferred one. The theory we consider consists of the
vector field $N^{A}$ minimally coupled to gravity with an action
of the form\footnote{The upper case Latin indices take the values
0, 1, 2, 3 and 5 while the Greek indices run from 0 to 3.}.
\begin{equation}
S=\int d^{5}x \sqrt{- ^{(5)}g}\left[\frac{1}{2k_{5}^{2}}
\left(^{(5)}R+{\cal L}_{N}\right)+{\cal L}_{m}\right],\label{eq1}
\end{equation}
where $k_{5}^{2}$ is a constant introduced for dimensional
considerations, ${\cal L}_{N}$ is the vector field Lagrangian
density while ${\cal L}_{m}$ denotes the Lagrangian density for
all the other matter fields. In order to preserve general
covariance, $N^{A}$ is taken to be a dynamical field. The
Lagrangian density for the vector field is written as
\begin{equation} {\cal
L}_{N}={K^{AB}}_{CD}\nabla_{A}N^{C}\nabla_{B}N^{D}+\lambda
(N^{A}N_{A}-\epsilon),\label{eq2}
\end{equation}
where
\begin{equation}
{K^{AB}}_{CD}=-\beta_{1}g^{AB}g_{CD}-
\beta_{2}\delta^{A}_{C}\delta^{B}_{D}-\beta_{3}\delta^{A}_{D}\delta^{B}_{C}.\label{eq3}
\end{equation}
Here, $\beta_{i}$ are dimensionless parameters, $\epsilon=-1$ or
$\epsilon=1$ depending on whether the extra dimension is
space-like or time-like respectively and  $\lambda$ is a Lagrange
multiplier. This is a slight simplification of the theory
introduced by Jacobson and Mattingly \cite{JM}, where we have
neglected a quartic self-interaction term of the form
$(N^{A}\nabla_{A}N^{B})(N^{c}\nabla_{c}N_{B})$ as has been done in
\cite{CL}. For their theory to be well-defined, the following
conditions should hold
\begin{equation}
\beta_{1}>0,\hspace{.5cm}\frac{(\beta_{1}+\beta_{2}+\beta_{3})}{\beta_{1}}\geq0,\hspace{.5cm}
\frac{(\beta_{1}+\beta_{2}+\beta_{3})}{\beta_{1}}\leq1,\hspace{.5cm}\beta_{1}+\beta_{3}\leq0.
\label{eqa}
\end{equation}
The first of these arises from the need for a positive-definite
Hamiltonian for the perturbation, the next two from demanding
non-tachyonic and subluminal propagation of the spin-0 degrees of
freedom respectively, and the last from insisting that gravity
waves propagate subluminally \cite{CL,EAL}. In the present work
however, we will not be relying on these constraints for our
analysis. We also define a current tensor ${J^{A}}_{C}$ via
\begin{equation}
{J^{A}}_{C}\equiv {K^{AB}}_{CD}\nabla_{B}N^{D}.\label{eq4}
\end{equation}
Note that the symmetry of ${K^{AB}}_{CD}$ means that
${J^{B}}_{D}={K^{AB}}_{CD}\nabla_{A}N^{C}$. With these definitions
the equation of motion obtained by varying the action with respect
to $ N^{A}$ is
\begin{equation}
\nabla_{A}J^{AB}=\lambda N^{B}.\label{eq5}
\end{equation}
The equation of motion for $\lambda$ enforces the fixed norm
constraint
\begin{equation}
g_{AB}N^{A}N^{B}=\epsilon, \hspace{1cm} \epsilon^{2}=1.\label{eq6}
\end{equation}
Choosing $\epsilon=1$ ensures that the vector will be time-like.
Multiplying both sides of (5) by $N_{B}$ and using (\ref{eq6}), we
find
\begin{equation}
\lambda=\epsilon N_{B}\nabla_{A}J^{AB}.\label{eq7}
\end{equation}
One may also project into a subspace orthogonal to $N^{A}$ by
acting the projection tensor ${P^{C}}_{B}=-\epsilon
N^{C}N_{B}+\delta^{C}_{B}$ on equation (\ref{eq5}) to obtain
\begin{equation}
\nabla_{A}J^{AC}-\epsilon N^{C}N_{B}\nabla
_{A}J^{AB}=0.\label{eq8}
\end{equation}
This equation determines the dynamics of $N^{A}$, subject to the
fixed-norm constraint.

In taking the variation, it is important to distinguish the
variables that are independent. Our dynamical degrees of freedom
are the inverse metric $g^{AB}$ and the contravariant vector field
$N^{A}$. Hence, the Einstein equations in the presence of both the
matter and vector fields in bulk space are \cite{EJM}
\begin{eqnarray}
^{(5)}G_{AB}=^{(5)\!\!\!}R_{AB}-\frac{1}{2}g_{AB}
{^{(5)}\!}R=k_{5}^2{^{(5)}}T_{AB},\label{eq9}
\end{eqnarray}
where
\begin{equation}
 {^{(5)}}T_{AB}=^{(5)\!\!}T^{(m)}_{AB}+\frac{1}{k^{2}_{5}}T^{(N)}_{AB}.\label{eqa.1}
\end{equation}
Here, $T_{AB}^{(m)}$ is the five-dimensional energy-momentum
tensor and the stress-energy $T_{AB}^{(N)}$ is considered to have
the following form \cite{CL,TJDM}
\begin{eqnarray}
T^{(N)}_{AB}&=&2\beta_{1}\left(\nabla_{A}N^{C}\nabla_{B}N_{C}-
\nabla^C
N_{A}\nabla_{C}N_{B}\right)-2\left[\nabla_C\left(N_{(A}{J^C}_{B)}\right)
+\nabla_{C}\left(N^C J_{(AB)}\right)\right.\nonumber\\ &-& \left.
\nabla_{c}\left(N_{(A} {J_{B)}}^{C}\right)\right]+ 2\epsilon
N_D\nabla_C J^{CD}N_A N_B +g_{AB}{\cal L}_{N}.\label{eq10}
\end{eqnarray}
Let us take the metric of the bulk space as
\begin{equation}
dS^{2}=g_{\mu\nu}(x^{\alpha},y)dx^{\mu}dx^{\nu}+\epsilon
\phi^{2}(x^{\alpha},y)dy^{2},\label{eq11}
\end{equation}
where $\phi$ is a scalar field and we have used signature $(+ - -
- \epsilon)$ everywhere. The Einstein equations (\ref{eq9})
contain the first and second derivatives of the metric with
respect to the extra coordinate. These can be expressed in terms
of geometrical tensors in $4D$. In the absence of off-diagonal
terms $(g_{5\mu}=0)$ the dimensional reduction of the
five-dimensional equations is particularly simple \cite{ponce},
\cite{JPL}. The usual assumption is that our space-time is
orthogonal to the extra dimension. Thus using equations
(\ref{eq6}) and (\ref{eq11}), we can introduce the normal unit
vector $N^{A}$ which is orthogonal to the hypersurfaces
$y=\mbox{const.}$ as
\begin{equation}
N^{A}=\frac{\delta^{A}_{5}}{\phi}, \hspace{1cm}
N_{A}=(0,0,0,0,\epsilon\phi).\label{eq12}
\end{equation}
The first partial derivatives can be written in terms of the
extrinsic curvature
\begin{equation}
K_{\mu\nu}=\frac{1}{2}{\cal
L}_{N}g_{\mu\nu}=\frac{1}{2\phi}\frac{\partial
g_{\mu\nu}}{\partial y}, \hspace{1cm}K_{A5}=0.\label{eq13}
\end{equation}
The second derivatives can be expressed in terms of the projection
$^{(5)}C_{\mu5\nu5}$ of the bulk Weyl tensor to 5$D$
\begin{equation}
^{(5)}C_{ABCD}=^{(5)\!\!\!}R_{ABCD}-\frac{2}{3}\left(^{(5)\!\!}R_{A[C}g_{D]B}-
^{(5)\!\!\!}R_{B[C}g_{D]A}\right)+\frac{1}{6}\left(^{(5)\!\!}R
g_{A[C}g_{D]B}\right).\label{eq14}
\end{equation}
The field equations (\ref{eq9}) can be split up into three parts.
Using the Gauss-Codazzi equations, the effective field equations
in $4D$ are
\begin{eqnarray}
^{(4)}G_{\mu\nu}&=&\frac{2}{3}k_{5}^{2}\left[^{(5)}T_{\mu\nu}+\left(^{(5)}T^{5}_{5}-\frac{1}{4}
(^{(5)}T)\right)g_{\mu\nu}\right]\nonumber\\
&-&\epsilon\left(K_{\mu\alpha}K^{\alpha}_{\nu}-K
K_{\mu\nu}\right)+\frac{\epsilon}{2}g_{\mu\nu}\left(K_{\alpha\beta}K^{\alpha\beta}-K^{2}\right)-\epsilon
E_{\mu\nu}, \label{eq15}
\end{eqnarray}
where
\begin{equation}
E_{\mu\nu}=^{(5)\!\!}C_{\mu A \nu
B}N^{A}N^{B}=-\frac{1}{\phi}\frac{\partial K_{\mu\nu}}{\partial
y}+K_{\mu\gamma}K^{\gamma}_{\nu}-\epsilon
\frac{\phi_{\mu;\nu}}{\phi}-\epsilon
\frac{k_{5}^{2}}{3}\left[^{(5)}T_{\mu\nu}+
\left(^{(5)}T^{5}_{5}-\frac{1}{2}(^{(5)}T)\right)g_{\mu\nu}\right].\label{eq16}
\end{equation}
Since the electric part of the Weyl tensor $E_{\mu\nu}$ is
traceless, the requirement $E^{\mu}_{\mu}=0$ gives the
inhomogeneous wave equation for $\phi$
\begin{equation}
\phi^{\mu}_{;\mu}=-\epsilon\frac{\partial K}{\partial y}-\phi
\left(\epsilon
K_{\alpha\beta}K^{\alpha\beta}+^{(5)\!\!}R^{5}_{5}\right),
\label{eq17}
\end{equation}
which is equivalent to $^{(5)}G_{55}=k_{5}^{2}\, ^{(5)}T_{55}$
from (\ref{eq9}). The remaining four equations are
\begin{equation}
D_{\mu}(K^{\mu}_{\nu}-\delta^{\mu}_{\nu}K)=
k^{2}_{(5)}\frac{^{(5)}T_{5\nu}}{\phi}.\label{eq18}
\end{equation}
In the above expressions, the covariant derivatives are calculated
with respect to $g_{\mu\nu}$, {\it i.e.}, $D g_{\mu\nu}=0$.
\section{Brane-world considerations}
With an eye on the brane-world scenario, it is assumed that the
five-dimensional energy-momentum tensor has the form
\begin{equation}
^{(5)}T^{(m)}_{AB}=\Lambda_{5}g_{AB},\label{eq19}
\end{equation}
where $\Lambda_{5}$ is the cosmological constant in the bulk. Now,
using equation (\ref{eq10}) we may calculate $^{(5)}T_{\mu\nu}$,
$^{(5)}T^{5}_{5}$ and $^{(5)}T$, obtaining
\begin{eqnarray}
^{(5)}T_{\mu\nu}&=&\frac{1}{k^{2}_{5}}\left[-4(\beta_{1}+\beta_{3})K_{\mu\gamma}K^{\gamma}_{\nu}+2(\beta_{1}+\beta_{3})K
K_{\mu\nu}+\beta_{2}g_{\mu\nu}K^{2}+\frac{2(\beta_{1}+\beta_{3})}{\phi}K_{\mu\nu,5}\nonumber
\right.\\
&+& \left. \frac{2\beta
{2}}{\phi}g_{\mu\nu}K_{,5}-(\beta_{1}+\beta_{3})g_{\mu\nu}K_{\alpha\beta}
K^{\alpha\beta}+2\epsilon\beta_{1}\frac{\phi_{,\mu}\phi_{,\nu}}{\phi^{2}}-
\epsilon\beta_{1}g_{\mu\nu}\frac{\phi_{,\alpha}\phi_{,}^{\alpha}}{\phi^{2}}\right]+\Lambda_{5}g_{\mu\nu},
\label{eq20}\\
^{(5)}T^{5}_{5}&=&\frac{1}{k^{2}_{5}}\left[(\beta_{1}+\beta_{3})K_{\alpha\beta}K^{\alpha\beta}+\beta_{2}K^{2}+
2\epsilon\beta_{1}g^{\mu\nu}\frac{\phi_{;\nu\mu}}{\phi}
-\epsilon\beta_{1}\frac{\phi_{,\alpha}\phi_{,}^{\alpha}}{\phi^{2}}\right]+\Lambda_{5},\label{eq21}\\
^{(5)}T&=&\frac{1}{k^{2}_{5}}\left[-3(\beta_{1}+\beta_{3})K_{\alpha\beta}K^{\alpha\beta}+2(\beta_{1}+\beta_{3})K^{2}+
5\beta_{2}K^{2}+\frac{2}{\phi}(\beta_{1}+\beta_{3})K_{,5}+\frac{8}{\phi}\beta_{2}
K_{,5} \nonumber \right.\\
&-& \left.
3\epsilon\beta_{1}\frac{\phi_{,\alpha}\phi_{,}^{\alpha}}{\phi^{2}}+
2\epsilon\beta_{1}g^{\mu\nu}\frac{\phi_{;\mu\nu}}{\phi}\right]+5\Lambda_{5}.
\label{eq22}
\end{eqnarray}
Let us now substitute equations (\ref{eq20}), (\ref{eq21}) and
(\ref{eq22}) in equation (\ref{eq15}) and use the following
equations
\begin{eqnarray}
\frac{1}{\phi}(K_{\mu\nu,5}-\frac{1}{4}g_{\mu\nu}K_{,5})&=&
-\frac{1}{2}\left(^{(4)}R_{\mu\nu}-\frac{1}{4}g_{\mu\nu}^{(4)}R\right)-
\left(\frac{\phi_{;\nu\mu}}{\phi}-\frac{1}{4}g_{\mu\nu}
\frac{\phi^{\alpha}_{\,\,\,;\alpha}}{\phi}\right)
\nonumber \\
&+&\frac{1}{2}\left(K_{\mu}^{\alpha}K_{\alpha\nu}+K
K_{\mu\nu}\right)+\frac{3}{8}g_{\mu\nu}\left(K_{\alpha\beta}K^{\alpha\beta}-\frac{1}{3}K^{2}\right)-
\frac{3}{2}E_{\mu\nu},\label{eq23}
\end{eqnarray}
\begin{eqnarray}
^{(4)}R&=&\epsilon K^{2}-\epsilon K_{\alpha\beta}K^{\alpha\beta}-2
\left(R^{5}_{5}-\frac{1}{2}g^{5}_{5}R\right)=\epsilon
K^{2}-\epsilon
K_{\alpha\beta}K^{\alpha\beta}-2k_{5}^{2} T^{5}_{5} =\epsilon K^{2} \nonumber\\
&-&\epsilon
K_{\alpha\beta}K^{\alpha\beta}-2k_{5}^{2}\Lambda_{5}-2(\beta_{1}+
\beta_{3})K_{\alpha\beta}K^{\alpha\beta}-2\beta_{2}K^{2}-4\epsilon
\beta_{1}g^{\alpha\beta}\frac{\phi_{;\alpha\beta}}{\phi}+
2\epsilon\beta_{1}\frac{\phi_{,\alpha}\phi_{,}^{\alpha}}{\phi^{2}},
\label{eq24}
\end{eqnarray}
where (\ref{eq23}) may be obtained from an equation for the
electric part of the Weyl tensor \cite{SMS} and (\ref{eq24}) can
be derived from the Gauss equation. Now, upon defining the
following new set of parameters
\begin{eqnarray}
\alpha_{1}&=&2(\beta_{1}+\beta_{3}), \hspace{1cm}
\alpha_{2}=\frac{2\epsilon(\beta_{1}+\beta_{2}+\beta_{3})}{3-2\epsilon
(\beta_{1}+4\beta_{2}+\beta_{3})}, \nonumber\\
\\
\alpha_{3}&=&\frac{\alpha_{1}(3+\epsilon-2\beta_{2})}{6}-\beta_{2},\hspace{1cm}
\alpha_{4}=\frac{\alpha_{1}(6+\epsilon+\alpha_{1})}{6},\hspace{1cm}
\alpha_{5}=\epsilon\beta_{1}, \nonumber\label{eq25}
\end{eqnarray}
equation (\ref{eq15}) becomes
\begin{eqnarray}
^{(4)}G_{\mu\nu}&=&\frac{k^{2}_{5}}{2}g_{\mu\nu}\Lambda_{5}-\frac{3(\epsilon
+\alpha_{1})}{(3+\alpha_{1})}(K_{\mu\gamma}K^{\gamma}_{\nu}-K
K_{\mu\nu})-\frac{3(\epsilon+\alpha_{3})}{2(3
+\alpha_{1})}g_{\mu\nu}K^{2} \nonumber \\
&+&\frac{3(\epsilon+\alpha_{4})}{2(3+\alpha_{1})}g_{\mu\nu}K_{\alpha\beta}K^{\alpha\beta}
+\left[\frac{\alpha_{1}(\alpha_{5}+\frac{1}{2})+
3\alpha_{5}}{(3+\alpha_{1})}\right]g_{\mu\nu}
\frac{\phi_{;\alpha}^{\alpha}}{\phi}-
\frac{\alpha_{5}(5+\alpha_{1})}{2(3+\alpha_{1})}
g_{\mu\nu}\frac{\phi_{,\alpha}\phi_{,}^{\alpha}}{\phi^{2}}
\nonumber \\
&-&\frac{2\alpha_{1}}{(3+\alpha_{1})}\frac{\phi_{;\nu\mu}}{\phi}+
\left[\frac{4\alpha_{5}}{(3+\alpha_{1})}\right]
\frac{\phi_{,\mu}\phi_{,\nu}}{\phi^{2}}-\frac{3(\epsilon+\alpha_{1})}{(3+\alpha_{1})}E_{\mu\nu}.
\label{eq26}
\end{eqnarray}
Note that $(3+\alpha_{1})$ is the coefficient of the
four-dimensional Einstein tensor. It therefore provides a relation
among the extrinsic curvature, the electric part of the Weyl
tensor and scalar field $\phi$ when $\alpha_{1}=-3$. We take
$\alpha_{1}\neq-3$ from hereon. In the spirit of the brane world
scenario, we assume $ Z_{2}$ symmetry about our brane, considered
to be a hypersurface $\Sigma$ at $y=0$, and write the
five-dimensional energy-momentum tensor $^{(5)}T^{(m)}_{AB}$ in
the resulting $Z_{2}$-symmetric brane universe as
\begin{equation}
^{(5)}T^{(m)}_{AB}=\Lambda_{5}g_{AB}+^{(5)}T^{(brane)}_{AB},\label{eq29}
\end{equation}
where $^{(5)}T^{(brane)}_{AB}$ is the energy-momentum tensor of
the matter on the brane with $^{(5)}T^{(brane)}_{AB} N^{A}=0$ and
\begin{equation}
^{(5)}T_{AB}^{(brane)}=\delta^{\mu}_{A}\delta^{\nu}_{B}\tau
_{\mu\nu}\frac{\delta(y)}{\phi}. \label{eq30}
\end{equation}
In order to obtain the Einstein equations on the brane, we need to
find an expression for the extrinsic curvature of $\Sigma$. For
this, we use equation (\ref{eq9}) and find
\begin{eqnarray}
^{(5)}R_{\mu\nu}&=&k_{5}^{2}\left[
\frac{-2}{3}\Lambda_{5}g_{\mu\nu}+\frac{\delta(y)}{\phi}\left(\tau_{\mu\nu}-
\frac{1}{3}g_{\mu\nu}\tau\right)\right]-4(\beta_{1}+\beta_{3})K_{\mu\gamma}
K ^{\gamma}_{\nu} +2(\beta_{1}+ \beta_{3})K K_{\mu\nu}
\nonumber\\
&-& \frac{2}{3}(\beta_{1}+\beta_{2}+\beta_{3})g_{\mu\nu}K^{2}
+\frac{2(\beta_{1}+\beta_{3})}{\phi}K_{\mu\nu,5}-\frac{2}{3}(\beta_{1}+\beta_{2}+\beta_{3})
g_{\mu\nu}\frac{K_{,5}}{\phi}\nonumber\\
&+&
2\epsilon\beta_{1}\frac{\phi_{,\mu}\phi_{,\nu}}{\phi^{2}}-\frac{2}{3}\epsilon
\beta_{1}g_{\mu\nu}\frac{\phi_{;\alpha}^{\alpha}}{\phi}.
\label{eq32}
\end{eqnarray}
On the other hand, for metric (\ref{eq11}) we have
\begin{equation}
^{(5)}R_{\mu\nu}=-\epsilon\frac{\partial}{\partial
y}\left(\frac{K_{\mu\nu}}{\phi}\right)+V_{\mu\nu}\label{eq33}
\end{equation}
where
\begin{equation}
V_{\mu\nu}=^{(4)}R_{\mu\nu}+\epsilon(2K_{\mu\gamma}K^{\gamma}_{\nu}-
K_{\mu\nu}K)-\frac{^{(5)}\nabla_{\nu}\phi_{;\mu}}{\phi},
\label{eq34}
\end{equation}
with
$^{(5)}\nabla_{\nu}\phi_{\mu}=\phi_{\mu,\nu}-\Gamma^{A}_{\mu\nu}\phi_{A}$.
We now substitute this into equation (\ref{eq32}) and integrate
across the brane noting that the metric is continuous. Although
the derivatives $\partial g_{\mu\nu}/\partial y$ and $ \partial
\phi/\partial y $ are discontinuous across $ \Sigma : y=0 $, we
make the usual physical assumption that they remain finite. Thus,
$ \lim_{\xi\rightarrow 0} \int^{\xi/2}_{-\xi/2} V_{\mu\nu}dy=0$
and using $Z_{2}$ symmetry and the set of parameters (\ref{eq25}),
we obtain
\begin{equation}
K_{\mu\nu}|_{\Sigma^{+}}=-K_{\mu\nu}|_{\Sigma^{-}}=-\frac{\epsilon
k^{2}_{5}}{2(1+\epsilon
\alpha_{1})}\left[\tau_{\mu\nu}-\frac{1}{3}g_{\mu\nu}(1+\alpha_{2})\tau\right].
\label{eq35}
\end{equation}
To avoid unreal singularities in equations (\ref{eq26}) and
(\ref{eq35}), it would be convenient to take $\alpha_{1}<-3$. Now,
from equation (\ref{eq18}) it follows that
\begin{equation}
(\tau^{\mu}_{\nu})_{;\mu}=-2(\epsilon+
\alpha_{1})\frac{^{(5)}T_{5\nu}}{\phi}-\alpha_{2}\tau_{,\nu},
\label{eq36}
\end{equation}
where
\begin{equation}
^{(5)}T_{5\nu}=\frac{2\epsilon}{k^{2}_{5}}(2\beta_{1}+\beta_{2}+2\beta_{3})K_{\nu}^{\mu}\phi_{,\mu}+
\frac{2\epsilon}{k^{2}_{5}}(2\beta_{2}-\beta_{1})K\phi_{,\nu}-
\frac{\epsilon\beta_{1}}{k^{2}_{5}}\frac{\phi_{,\nu5}}{\phi}.
\label{eq37}
\end{equation}
Thus, the energy-momentum tensor $ \tau_{\mu\nu}$ is not conserved
on the brane and represents the total vacuum plus matter
energy-momentum. It is usually separated in two parts,
\begin{equation}
\tau_{\mu\nu}=\sigma g_{\mu\nu}+T_{\mu\nu}, \label{eq38}
\end{equation}
where $ \sigma$ is the tension of the brane in $ 5D $, which is
interpreted as the vacuum energy of the brane world and
$T_{\mu\nu}$ represents the energy-momentum tensor of ordinary
matter in $4D$. Using equations (\ref{eq35}) and (\ref{eq38}) and
defining the following set of parameters
\begin{eqnarray}
\alpha_{6}&=&\frac{\alpha_{1}(1-2\alpha_{2})-2\epsilon\alpha_{2}}{3},
\ \nonumber\\
\\
\alpha_{7}&=&\frac{(\epsilon+\alpha_{1})(\alpha_{2}+\alpha_{2}^{2})}{3}+\frac{(\alpha_{4}-3\epsilon-4\alpha_{3})
(\alpha_{2}+2\alpha_{2}^{2})}{9}-\frac{(\alpha_{3}+2\alpha_{4})}{18},
\nonumber\label{eq39}
\end{eqnarray}
we obtain the Einstein field equations with an effective
energy-momentum tensor in $4D$ as
\begin{eqnarray}
^{(4)}G_{\mu\nu}&=&\Lambda_{4}g_{\mu\nu}+8 \pi G
T_{\mu\nu}+k^{4}_{5}\Pi_{\mu\nu}-
\frac{3(\epsilon+\alpha_{1})}{(3+\alpha_{1})}E_{\mu\nu}+\left[\frac{\alpha_{1}(\alpha_{5}+\frac{1}{2})+
3\alpha_{5}}{(3+\alpha_{1})}\right]g_{\mu\nu}
\frac{\phi_{;\alpha}^{\alpha}}{\phi}\nonumber \\&-&
\frac{\alpha_{5}(5+\alpha_{1})}{2(3+\alpha_{1})}
g_{\mu\nu}\frac{\phi_{,\alpha}\phi_{,}^{\alpha}}{\phi^{2}}
-\frac{2\alpha_{1}}{(3+\alpha_{1})}\frac{\phi_{;\nu\mu}}{\phi}+
\left[\frac{4\alpha_{5}}{(3+\alpha_{1})}\right]
\frac{\phi_{,\mu}\phi_{,\nu}}{\phi^{2}} , \label{eq40}
\end{eqnarray}
where
\begin{equation}
\Lambda_{4}=\frac{k^{2}_{5}}{2}\Lambda_{5}+\left[\frac{-\epsilon
k_{5}^{4}+3k_{5}^{4}(-\alpha_{1}+4\alpha_{6}+16\alpha_{7}+2\alpha_{4})
}{4(3+\alpha_{1}) (1+\epsilon\alpha_{1})^{2}}\right]\sigma^{2},
\label{eq41}
\end{equation}
\begin{equation}
8 \pi G=\left[\frac{-2\epsilon
k_{5}^{4}+3k_{5}^{4}(-2\alpha_{1}+4\alpha_{6})}
{4(3+\alpha_{1})(1+\epsilon\alpha_{1})^{2}}\right]\sigma,
\label{eq42}
\end{equation}
and
\begin{eqnarray}
\Pi_{\mu\nu}&=&\frac{3}{4(3+\alpha_{1})(1+\epsilon\alpha_{1})^{2}}
\left[-(\epsilon+\alpha_{1})T_{\mu\gamma}T^{\gamma}_{\nu}+(\frac{\epsilon}{3}+\alpha_{6})T
T_{\mu\nu}\right. \nonumber\\
&-&\left.
\left(\frac{\epsilon}{6}-\alpha_{7}\right)g_{\mu\nu}T^{2}+\frac{(\epsilon+\alpha_{4})}{2}g_{\mu\nu}
T_{\alpha\beta}T^{\alpha\beta}\right]+\left[\frac{3(\alpha_{6}+8\alpha_{7}+\alpha_{4})}
{4(3+\alpha_{1})(1+\epsilon\alpha_{1})^{2}}\right]g_{\mu\nu}\sigma
T. \label{eq43}
\end{eqnarray}
All these $4D$ quantities have to be evaluated in the limit $
y\rightarrow 0^{+}$. They give a working definition of the
fundamental quantities $ \Lambda_{4}$ and $G$ and contain
higher-dimensional modifications to general relativity. As
expected, switching off the effects of Lorentz violations
$(\alpha_i=0)$ in these equations results in expressions one
usually obtains in the brane-worlds models.
\section{Cosmological implications}
Assuming a perfect fluid configuration on the brane, the
energy-momentum tensor is written as
\begin{equation}
T_{\mu\nu}=(\rho+p)u_{\mu}u_{\nu}-pg_{\mu\nu}, \label{eq44}
\end{equation}
where $\textbf{u}$, $\rho$ and $p$ are the unit velocity, energy
density and pressure of the matter fluid respectively. We will
also assume a linear isothermal equation of state for the fluid
\begin{equation}
p=(\gamma-1)\rho,\hspace{10mm} 1\leq\gamma\leq 2. \label{eq45}
\end{equation}
The weak energy condition \cite{HE} imposes the restriction
$\rho\geq0$. Now, it is useful to decompose $^{(5)}E_{\mu\nu}$
with respect to any time-like observers
$\textbf{u}\hspace{1mm}(u^{\mu}u_{\mu}=1)$ into a scalar part,
$U$, a vector part, $Q_{\mu}$, and a tensorial part $P_{\mu\nu}$
\cite{RM}, that is
\begin{equation}
^{(5)}E_{\mu\nu}=-\left(\frac{k_{5}}{k_{4}}\right)^{4}\left[\left(u_{\mu}u_{\nu}+
\frac{1}{3}h_{\mu\nu}\right)U+2u_{(\mu}Q_{\nu)}+P_{\mu\nu}\right],
\label{eq46}
\end{equation}
where the following properties hold
\begin{equation}
Q_{\mu}u^{\mu}=0,\hspace{1cm} P_{\mu\nu}=P_{\nu\mu},\hspace{1cm}
P^{\mu}_{\mu}=0,\hspace{1cm} P_{\mu\nu}u^{\nu}=0. \label{eq47}
\end{equation}
Here, $Q_{\mu}$ is a spatial vector and $P_{\mu\nu}$ is a spatial,
symmetric and trace-free tensor. The scalar term has the same form
as the energy-momentum tensor of a radiation perfect fluid and for
this reason $U$ is called the dark energy density \cite{RM}. From
the modified Einstein equations (\ref{eq40}) and equations
(\ref{eq36}) and (\ref{eq38}) we may  obtain a constraint on
$^{(5)}E_{\mu\nu}$ as has been done in \cite{CS}.

In this paper we deal with non-tilted homogeneous cosmological
models on the brane, {\it i.e.} we are assuming that the fluid
velocity is orthogonal to the hypersurfaces  of homogeneity. Also,
we may consider $\phi(x^{\alpha}, y)=\phi(t)$. It is then
convenient to make the splitting (\ref{eq46}) with respect to the
fluid velocity. In particular, we will consider  FLRW models with
the metric tensor given by
\begin{equation}
dS^{2}=dt^{2}-a(t)^{2}\left[dr^{2}+{\cal
Q}^{2}_{k}(r)(d\theta^{2}+\sin^{2}\theta d\varphi^{2})\right],
\label{eq48}
\end{equation}
where ${\cal Q}_{k}(r)=\sin(r),r,\sinh(r)$ corresponds to
$k=1,0,-1$ respectively and $a(t)$ is the scale factor, with the
fluid velocity given by $\textbf{u}=\frac{\partial}{\partial t}$.
Taking into account these assumptions and the effective Einstein's
equations (\ref{eq40}), the consequences of having a FLRW universe
on the brane are
\begin{equation}
Q_{\mu}=P_{\mu\nu}=0, \hspace{1cm} U=U(t). \label{eq50}
\end{equation}
The brane evolution can now be studied by determining the
generalized Friedmann and Raychaudhuri equations in the presence
of the vector field in the bulk. These follow from the
Gauss-Codazzi equations and the time-like part of the trace of the
Ricci identities applied to the time-like congruence $u_{\mu}$.
Defining the parameters
\begin{eqnarray}
\alpha_{8}&=&(1+3(\gamma-1)^{2})(2\alpha_{4}-\alpha_{1})+(4-3\gamma)^{2}(\alpha_{6}+4\alpha_{7}),\nonumber\\
\alpha_{9}&=&(4-3\gamma)(\alpha_{6}+8\alpha_{7}+\alpha_{4}),\\
\alpha_{10}&=&\frac{(1+3(\gamma-1)^{2})\alpha_{4}}{2}
+(4-3\gamma)^{2}\alpha_{7}+(4-3\gamma)\alpha_{6}-\alpha_{1},\nonumber
\label{eq51}
\end{eqnarray}
these equations can be written as
\begin{eqnarray}
H^{2}&=&\frac{1}{3}\Lambda_{4}+\frac{1}{3}\left[1+\frac{\alpha_{9}}{(-\frac{2\epsilon}{3}-2\alpha_{1}+4\alpha_{6})}\right]
k_{4}^{2}\rho+\frac{1}{3}\left[\frac{-\frac{\epsilon}{3}+\alpha_{10}}{(-\frac{2\epsilon}{3}-2\alpha_{1}+4\alpha_{6})\sigma}
\right]k_{4}^{2}\rho^{2}
+\left(\frac{k_{5}}{k_{4}}\right)^{4}\left(\frac{\epsilon+\alpha_{1}}{3+\alpha_{1}}\right)U(t)
\nonumber\\
&+&\left[\frac{\alpha_{1}(\alpha_{5}-
\frac{3}{2})+3\alpha_{5}}{3(3+\alpha_{1})}\right]\frac{\ddot{\phi}}{\phi}+\left
[\frac{\alpha_{5}(3-\alpha_{1})}{6(3+\alpha_{1})}\right]
\left(\frac{\dot{\phi}}{\phi}\right)^{2}-\frac{k}{a(t)^{2}},
\label{eq52}
\end{eqnarray}
and
\begin{eqnarray}
\dot{H}&=&-\frac{\gamma}{2}k_{4}^{2}\rho+
\left[\frac{(2\epsilon\gamma+\alpha_{8}-4\alpha_{10})}{6(-\frac{2\epsilon}{3}-2\alpha_{1}+4\alpha_{6})
\sigma}\right]k_{4}^{2}\rho^{2}
-\left(\frac{k_{5}}{k_{4}}\right)^{4}\left[\frac{2(\epsilon+\alpha_{1})}{3+\alpha_{1}}\right]U(t)
\nonumber\\
&+&\left[\frac{\alpha_{1}}{(3+\alpha_{1})}\right]\frac{\ddot{\phi}}{\phi}
-\left[\frac{2\alpha_{5}}{(3+\alpha_{1})}\right]
\left(\frac{\dot{\phi}}{\phi}\right)^{2}+\frac{k}{a(t)^{2}},
\label{eq53}
\end{eqnarray}
allowing us to examine the evolution of the brane without using
any particular solution of the five-dimensional field equations.
It is worth noting again that if the effects of Lorentz violations
are ignored $(\alpha_i=0)$, the above equations reduce to the
usual Friedmann and Raychaudhuri equations in the brane-world
scenarios \cite{CS}.
\subsection{Variable vacuum energy}
In the above equations $G$ and $\Lambda_{4}$ are usually assumed
to be truly constants. In this section we show how equations
(\ref{eq52}) and (\ref{eq53}) should be modified as to incorporate
the variation of these fundamental physical ``constants,''
matching observational predictions. We can make a simplification
without loss of generality by setting $\beta_{1}=2\beta_{2}$ and
$\beta_{3}=\frac{-5}{2}\beta_{2}$. Also noting that
$\alpha_{1}<-3$, we obtain $\beta_{2}>3$. From (\ref{eq36}),
(\ref{eq37}) and (\ref{eq38}) it follows that
\begin{equation}
\sigma_{,\nu}+T^{\mu}_{\nu;\mu}=-\alpha_{2}(4\sigma+T)_{,\nu},
\label{54}
\end{equation}
where $\alpha_{2}=\frac{\epsilon\beta_{2}}{3-7\epsilon\beta_{2}}$.
For the perfect fluid, equation (\ref{eq44}), this is equivalent
to
\begin{equation}
\left(1+\alpha_{2}(4-3\gamma)\right)\dot{\rho}+(\rho+p)\Theta=-(1+4\alpha_{2})\dot{\sigma},
\label{55}
\end{equation}
and
\begin{equation}
(\rho+p)a_{\nu}+\left[(\gamma-1)-\alpha_{2}(4-3\gamma)\right]\rho_{,\lambda}h^{\lambda}_{\nu}
=(1+4\alpha_{2})\sigma_{,\lambda}h^{\lambda}_{\nu},
\label{56}
\end{equation}
where $\Theta=u^{\mu}_{\,\,\,;\mu}$ is the usual expression for
expansion, $a_{\nu}=u_{\nu;\lambda}u^{\lambda}$ is the
acceleration and $h_{\mu\nu}=u_{\mu}u_{\nu}-g_{\mu\nu}$ is the
projection onto the spatial surfaces orthogonal to $u_{\mu}$. In
homogeneous cosmological models, equation (\ref{56}) becomes
redundant and only equation (\ref{55}) is relevant. For the case
of constant vacuum energy $\sigma$ and equation of state
$p=(\gamma-1)\rho$, it yields the familiar relation between the
matter energy density and expansion factor $a$
\begin{equation}
\rho=\rho_{0}
\left(\frac{a_{0}}{a}\right)^{\frac{3\gamma}{1+\alpha_{2}(4-3\gamma)}}.\label{eq57}
\end{equation}
For the case where the vacuum energy is not constant we need some
additional assumption. For example, $\sigma$ may be a function of
the scale factor, $\sigma=\sigma(a)$. However, we have so far no
theoretical observational arguments for the evolution of $\sigma$
in time.
\subsection{Time evolution of the universe}
The time variation of $G$ is usually written as
$\left(\frac{\dot{G}}{G}\right)=g H$, where $g$ is a dimensionless
parameter. Nucleosynthesis and the abundance of various elements
are used to put constraints on $g$. The present observational
upper bound is $|g|\leq 0.1$ \cite{JU, BMN}. In what follows we
assume that $g$ is constant. Since $G\sim\sigma$ and
$H=\frac{\dot{a}}{a}$ we have $\sigma(a)=f_{0}a^{g}$, where
$f_{0}$ is a constant of integration. From equation (\ref{55}) we
thus find
\begin{equation}
[1+\alpha_{2}(4-3\gamma)]\dot{\rho}+3\gamma\rho\frac{\dot{a}}{a}=
-(1+4\alpha_{2})f_{0}g a^{(g-1)}\dot{a}. \label{58}
\end{equation}
First, we consider the case where $\rho$ can be expressed in a way
similar to (\ref{eq57}), {\it i.e.} as a power function of $a$. We
therefore find $$g=\frac{-3\gamma D}{D+f_{0}+\alpha_{2}(4D-3\gamma
D+4f_{0})},$$ and
\begin{equation}
\rho=D a^{-\left[\frac{3\gamma D}{D+f_{0}+\alpha_{2}(4D-3\gamma
D+4f_{0})}\right]}, \label{59}
\end{equation}
where $D$ is a positive constant. In order to simplify the
notation we set $f_{0}=F_{0}D$ and
\begin{equation}
\frac{\gamma}{1+F_{0}+\alpha_{2}(4-3\gamma+4F_{0})}=\zeta+1,
\label{60}
\end{equation}
which gives
$$\zeta=\frac{\gamma-F_{0}-1-\alpha_{2}(4-3\gamma+4F_{0})}{1+F_{0}+\alpha_{2}(4-3\gamma+4F_{0})}.$$
With this notation and substituting $\alpha_{2}$, we have
\begin{equation}
\rho=\frac{D}{a^{3(\zeta+1)}} \hspace{0.5cm}
\mbox{and}\hspace{0.5cm}
\sigma=\left[\frac{(3-7\epsilon\beta_{2})D\gamma-3D(\zeta+1)
(1-\epsilon\beta_{2}-\epsilon\gamma\beta_{2})}{3(\zeta+1)(1-\epsilon\beta_{2})
}\right]a^{-3(\zeta+1)}. \label{eq61}
\end{equation}
Note that one may take $D=\rho_{0}a_{0}^{3(\zeta+1)}$ and that
$F_{0}\neq 0$
$\left(\zeta\neq\frac{\gamma-(1+\alpha_{2}(4-3\gamma))}{(1+\alpha_{2}(4-3\gamma))}\right)$,
otherwise $G=0$. Here, equations (\ref{eq41}) and (\ref{eq42}) can
be written as follows
\begin{eqnarray}
\Lambda_{4}&=&\frac{k_{5}^{2}}{2}\Lambda_{5}+\left[\frac{-\frac{3}{4}\epsilon+\frac{3}{2}\beta_{2}-\frac{1}{2}
\beta_{2}^{2}-\frac{65}{6}\epsilon\beta_{2}+\frac{29}{3}\epsilon\beta_{2}^{2}+\frac{17}{36}\epsilon\beta_{2}^{3}}
{(3-\beta_{2})(\epsilon-\beta_{2})^{2}(3-7\epsilon\beta_{2})}\right]
\nonumber\\&\times&\left[\frac{(3-7\epsilon\beta_{2})D\gamma-3D(\zeta+1)
(1-\epsilon\beta_{2}-\epsilon\gamma\beta_{2})}{3(\zeta+1)(1-\epsilon\beta_{2})
}\right]^{2}k_{5}^{4}a^{-6(\zeta+1)},\label{eqb1}\\
8\pi G&=&\left[\frac{-3\epsilon
k_{5}^{4}}{2(3-\beta_{2})(3-7\epsilon\beta_{2})}\right]
\left[\frac{(3-7\epsilon\beta_{2})D\gamma-3D(\zeta+1)(1-\epsilon\beta_{2}-\epsilon\gamma\beta_{2})}
{3(\zeta+1)(1-\epsilon\beta_{2}) }\right]a^{-3(\zeta+1)} .
\label{eqb2}
\end{eqnarray}
 We also have
\begin{equation}
\frac{\dot{G}}{G}=-3(\zeta+1)H. \label{eq62}
\end{equation}
Before going any further, we should be aware of the observational
bounds on $\zeta$. The lower bound comes from the obvious
requirement $\frac{d\rho}{da}<0$, while the upper bound comes from
the observation that $|g|\leq 0.1$. Thus,
\begin{equation}
-1<\zeta\leq-0.966, \hspace{1cm}\zeta=-0.983\pm0.016.\label{eq63}
\end{equation}
We can see that in general $\zeta$ is related to $q$ and
$\Omega_{\rho}$, the deceleration and density parameters
respectively. Another point of interest is that if one assumes a
constant bulk cosmological constant, equations (\ref{eq41}) and
(\ref{eq42}) imply $\dot{\Lambda}_4\sim G\dot{G}$, a relation
frequently encountered in brane theories.

The next step would be to obtain the evolution equation for $a$.
In the standard cosmological models, we can set
$\phi=\phi_{0}(\frac{a}{a_{0}})^{r}$. The physical condition
$\tau^{0}_{0}=(\sigma+\rho)>0$ then puts a lower limit on $r$,
namely $r>1$ \cite{ponce}. Now, substituting equations
(\ref{eqb1}) and (\ref{eqb2}) into
\begin{eqnarray}
(2-\frac{2}{3}\epsilon r^{2}\beta_{2})H^{2}+(1-\frac{4}{3}\epsilon
r
\beta_{2})\dot{H}&=&\frac{2}{3}\Lambda_{4}+\frac{2}{3}\left[\frac{4-3\gamma}{4}+\frac{\alpha_{9}}
{(-\frac{2\epsilon}{3}-2\alpha_{1}+4\alpha_{6})}\right]
k_{4}^{2}\rho\nonumber\\&+&
\left[\frac{-\frac{2}{3}(2-3\gamma)\epsilon+\alpha_{8}}{6(-\frac{2\epsilon}{3}-2\alpha_{1}+4\alpha_{6})
\sigma}\right]k_{4}^{2}\rho^{2}-\frac{k}{a^{2}}, \label{64}
\end{eqnarray}
which is obtained by combining equations (\ref{eq52}) and
(\ref{eq53}), and introducing the quantities
\begin{eqnarray}
e^{x}&=&a^{3(\zeta+1)},\nonumber\\
s_{1}&=&\left[\frac{6D^{2}(\zeta+1)}{(3-4\epsilon r
\beta_{2})}\right]\left[\frac{-\frac{3}{4}\epsilon+\frac{3}{2}\beta_{2}-\frac{1}{2}
\beta_{2}^{2}-\frac{65}{6}\epsilon\beta_{2}+\frac{29}{3}\epsilon\beta_{2}^{2}+\frac{17}{36}\epsilon\beta_{2}^{3}}
{(3-\beta_{2})(\epsilon-\beta_{2})^{2}(3-7\epsilon\beta_{2})}\right]
\nonumber\\&\times&\left[\frac{(3-7\epsilon\beta_{2})\gamma-3(\zeta+1)
(1-\epsilon\beta_{2}-\epsilon\gamma\beta_{2})}{3(\zeta+1)(1-\epsilon\beta_{2})
}\right]^{2}k_{5}^{4},\nonumber\\
s_{2}&=&\left[\frac{(4-3\gamma)D^{2}}{3-4\epsilon
r\beta_{2}}\right]\left[\frac{-\frac{3}{4}\epsilon+\frac{3}{2}\beta_{2}-\frac{1}{2}
\beta_{2}^{2}-\frac{65}{6}\epsilon\beta_{2}+\frac{29}{3}\epsilon\beta_{2}^{2}+\frac{17}{36}\epsilon\beta_{2}^{3}}
{(3-\beta_{2})(\epsilon-\beta_{2})^{2}(3-7\epsilon\beta_{2})}\right]
\nonumber\\&\times&\left[\frac{(3-7\epsilon\beta_{2})\gamma-3(\zeta+1)
(1-\epsilon\beta_{2}-\epsilon\gamma\beta_{2})}{(1-\epsilon\beta_{2})
}\right]k_{5}^{4},\nonumber\\
s_{3}&=&\left[\frac{D^{2}(\zeta+1)(4-3\gamma)^{2}}{3-4\epsilon r
\beta_{2}}\right]\left[\frac{3+9(\gamma-1)^{2}}{8(\epsilon-\beta_{2})(4-3\gamma)^{2}}+
\frac{9(\epsilon-\beta_{2})(1-3\epsilon\beta_{2})+2\epsilon\beta_{2}^{2}(1+\beta_{2})}
{8(3-7\epsilon\beta_{2})(3-\beta_{2})(1-\epsilon\beta_{2})^{2}}\right.\nonumber\\
&+& \left.\frac{\beta_{2}^{2}}
{12(3-\beta_{2})(1-\epsilon\beta_{2})^{2}}
+\frac{(-6+\beta_{2})}{8(3-\beta_{2})
(\epsilon-\beta_{2})}+\frac{\beta_{2}}{24(1-\epsilon\beta_{2})^{2}}\right]k_{5}^{4},\label{eq65}
\end{eqnarray}
with
\begin{eqnarray}
s&=&s_{1}+s_{2}+s_{3},\nonumber
\end{eqnarray}
we obtain the following equation
\begin{equation}
\ddot{x}+\left[\frac{6-2\epsilon
r^{2}\beta_{2}}{3(\zeta+1)(3-4\epsilon
r\beta_{2})}\right]\dot{x}^{2}=\left[\frac{3(\zeta+1)}{3-4\epsilon
r\beta_{2}}\right]k_{5}^{2}\Lambda_{5}+s
e^{-2x}-\left[\frac{9(\zeta+1)}{3-4\epsilon r \beta_{2}}\right]k
e^{\frac{-2x}{3(\zeta+1)}}. \label{eq66}
\end{equation}
Let us concentrate on the evolution of the universe at the present
epoch. In this case the exponential terms become small and could
be ignored. As it can be seen, the coefficient of $\dot{x}^{2}$ is
always positive. Therefore, the evolution of the universe not only
depends on the bulk cosmological constant being negative, zero or
positive but also depends on the extra dimension being timelike or
spacelike. We study these cases separately in what follows.
\subsubsection{Anti-de Sitter bulk, $\epsilon >0$}
This case is important since it corresponds to the brane-world
scenarios where our universe is identified with a singular
hypersurface embedded in an AdS$_5$ bulk. The evolution of the
scale factor is given by
\begin{equation}
a=a_{0}\left(\frac{\cosh\theta}{\cosh\theta_{0}}\right)^{\frac{1}{3(\zeta+1)E_{1}}}\label{eq67}
\end{equation}
where
\begin{eqnarray}
\theta=
\sqrt{E_{1}E_{2}}t^\prime+\tanh^{-1}\left(3(\zeta+1)H_{0}
\sqrt{\frac{E_{1}}{E_{2}}}\right),
\hspace{1cm}t^\prime=t-t_{0},\\
E_{1}=\left[\frac{6-2\epsilon
r^{2}\beta_{2}}{3(\zeta+1)(3-4\epsilon
r\beta_{2})}\right],\hspace{1cm}
E_{2}=\left[\frac{3(\zeta+1)}{3-4\epsilon
r\beta_{2}}\right]k_{5}^{2}\Lambda_{5}. \label{eq68}
\end{eqnarray}
The zero subscript denotes the measurement of the quantity at the
present epoch and $H$ is the Hubble parameter. Simply, using
equations (\ref{eqb1}) and (\ref{eqb2}), one can obtain the time
varying forms of $\Lambda_{4}$ and $G$. The deceleration parameter
$q$ defined by $q=\frac{-\ddot{a}a}{\dot{a}^{2}}$ becomes
\begin{equation}
q=-\left[\frac{3(\zeta+1)E_{1}(1-\tanh^{2}\theta)+\tanh^{2}\theta}
{\tanh^{2}\theta}\right],\label{eq69}
\end{equation}
which is always negative. Therefore, according to the
approximation made to solve equation (\ref{eq66}), one has late
time accelerated expansion of the universe. We note that for a
spacelike extra dimension the time-evolution of the brane in a
dS$_{5}$ bulk is again given by (\ref{eq67}).

We note that a non-vanishing cosmological constant in the bulk
induces a natural time scale in $4D$, defined as
$\tau_s=\sqrt{3/\tilde{\Lambda}}$ where
$\tilde{\Lambda}\equiv\frac{k_{5}^{2}\Lambda_{5}}{2(1-\frac{\epsilon}{3}r^{2}\beta_{2})}
$ \cite{ponce}. The influence of  Lorentz violation in the value
of $\tilde{\Lambda}$ is embodied in the factor $\beta_2$ and the
signature of the extra dimension. In the absence of Lorentz
violation, $\beta_2=0$ and one gets the usual expression. The late
time behavior of solutions (\ref{eq67}) becomes identical to the
one in the de Sitter solution
\begin{equation}
a(t^\prime)\sim\exp\left(\frac{t^\prime}{\tau_{s}}\right),\label{eqr}
\end{equation}
as $t^\prime\gg\tau_{s}$. This exponential behavior is not a
consequence of the false-vacuum equation of state $p= -\rho $ as
in inflation since $\gamma\neq 0$ here. The reason for this is
that the assumption $\frac{\dot{G}}{G}=g H$ is equivalent to the
requirement that $\rho$ and $\sigma$ form a combined fluid with
energy density $\tilde{\rho}=\rho+\sigma$ and pressure
$\tilde{p}=\zeta\tilde{\rho}$. Then, the observational constraint
$-1<\zeta\leq-0.966$ implies that the combined fluid behaves
nearly like a cosmological constant, which dominates at late times
and thus producing inflation.
\subsubsection{Bulk with $\Lambda_{5}=0$}
In this case, the expansion factor is given by
\begin{equation}
a=a_{0}\left[3(\zeta+1)H_{0}E_{1}t^\prime+1\right]^{\frac{1}{3(\zeta+1)E_{1}}}.
\label{eq70}
\end{equation}
One then obtains the deceleration parameter $q$ as follows
\begin{equation}
q=-\left[\frac{-3+2\epsilon r (r-2) \beta_{2}}{3-4\epsilon
r\beta_{2}}\right]. \label{eqd}
\end{equation}
To be in agreement with observations, the deceleration parameter
$q$ must be negative. This leads to the following condition
\begin{equation}
r(r-2)<\frac{3\epsilon}{2\beta_{2}},\label{eqdd}
\end{equation}
making the functional form of $\phi$ somewhat restricted.
\subsubsection{de Sitter bulk, $\epsilon >0$}
Although in brane-world theories our universe is embedded in a
higher-dimensional space with negative cosmological constant, the
solutions to the evolution equation depends analytically on the
sign of $E_{2}$, allowing us to consider the negative values of
$E_{2}$ which would include a positive cosmological constant,
provided the extra dimension is taken to be timelike. The
evolution of the scale factor is given by
\begin{equation}
a=a_{0}\left(\frac{\cos\varphi}{\cos\varphi_{0}}\right)^{\frac{1}{3(\zeta+1)E_{1}}},
\label{eq71}
\end{equation}
where
\begin{equation}
\varphi=-\sqrt{-E_{1}E_{2}}t^\prime+\tan^{-1}\left(3(\zeta+1)H_{0}\sqrt{\frac{-E_{1}}{E_{2}}}\right),
\label{eq72}
\end{equation}
and the deceleration parameter is found to be
\begin{equation}
q=-\left[\frac{3(\zeta+1)E_{1}(1+\tan^{2}\varphi)+\tan^{2}\varphi}{\tan^{2}\varphi}\right].
\label{eq73}
\end{equation}
In this case, $q$ is always negative, leading to an accelerated
expanding universe. Note that for $\epsilon <0$, the brane
evolution in an AdS$_{5}$ bulk is again given by (\ref{eq71}).
\section{Conclusions}
In this paper we have studied a brane-world scenario whereupon the
idea of Lorentz violation, introduced by specifying a preferred
frame through the introduction of a  dynamical vector field normal
to our brane has been considered. Such a normal vector was,
however, assumed to be decoupled from the matter fields since such
fields were assumed to be confined to the brane. The Einstein
field equations were obtained on the brane using the SMS
formalism, modified by the additional vector field. To study the
cosmological implication of our brane-world, Lorentz violating
model, we focused  attention on a Robertson-Walker background. The
ensuing Friedmann and Raychaudhuri equations were analyzed in
various cases where the bulk cosmological constant was negative or
positive together with the signature of the extra dimension being
timelike or spacelike. We found that an accelerated expanding
universe results in all these cases. For a zero bulk cosmological
constant, acceleration is possible only if a certain constraint on
$r$ is satisfied. Also, it was shown that local Lorentz violation
in the bulk allows for the construction of models in which the
vacuum energy $\sigma$, the gravitational coupling $G$ and the
cosmological term $\Lambda_{4}$ are variable.

\end{document}